\documentstyle[12pt,aaspp4]{article}
\input psfig.tex
\slugcomment{To appear in the Astrophysical Journal}

\begin{document}
\title{THE BEAMING PATTERN OF DOPPLER BOOSTED THERMAL ANNIHILATION RADIATION: 
APPLICATION TO MEV BLAZARS}
\author{Jeffrey G. Skibo and Charles D. Dermer}
\affil{E. O. Hulburt Center for Space Research, Code 7653, Naval Research
Laboratory, Washington, DC 20375-5352}

\and
\author {Reinhard Schlickeiser}
\affil{Max-Planck-Institut f\"ur Radioastronomie, Postfach 20 24, 53010 Bonn,
Germany}

\begin{abstract} 
The beaming pattern of thermal annihilation radiation is broader than the
beaming pattern produced by isotropic nonthermal electrons and positrons in the
jets of radio-emitting active galactic nuclei which Compton scatter photons
from an external isotropic radiation field.  Thus blueshifted thermal
annihilation radiation can provide the dominant contribution to the high-energy
radiation spectrum at observing angles $\theta \gtrsim 1/\Gamma$, where
$\Gamma$ is the bulk Lorentz factor of the outflowing plasma. This effect may
account for the spectral features of MeV blazars discovered with the Compton
Telescope on the {\it Compton Gamma Ray Observatory}.  Coordinated  gamma-ray
observations of annihilation line radiation to infer Doppler factors and VLBI
radio observations to measure transverse angular speeds of outflowing plasma
blobs can be used to determine the Hubble constant. 
\end{abstract}

\keywords {active galactic nuclei --- cosmology: distance scale --- galaxies: 
radio --- gamma rays: galaxies --- gamma rays: theory}

\section{Introduction}

Observations made with the Compton Telescope (Comptel) on the {\it Compton
Gamma-Ray  Observatory} ({\it CGRO}) reveal the existence of a class of
gamma-ray blazars displaying spectral energy distributions peaking in a rather
narrow energy band centered at a few MeV (Bloemen et al.
\markcite{Bloem95}1995; Blom et al. \markcite{Blom95}1995).  By combining data
from 19 September - 3 October 1991 and 27 December 1991 - 10 January 1992,
Bloemen et al. (\markcite{Bloem95}1995) discovered the first ``MeV blazar" GRO
J0516-609, which shows a 3-4$\sigma$ peak in its $\nu F_\nu$ spectrum in the 1
- 10 MeV energy interval.  Only upper limits are measured with Comptel in the
0.75-1 MeV and 10-30 MeV range, although contemporaneous observations with the
Energetic Gamma Ray Experiment Telescope (EGRET) on {\it CGRO} indicate the
presence of a weak $3\sigma$ excess at photon energies $E>100$ MeV.  The flat
spectrum quasars PKS 0506-612 and PKS 0522-611 are
both within the 95\% Comptel contours of GRO J0516-609, but the former source
is the preferred identification due to the more precise EGRET localization,
which places only PKS 0506-612 within its 95\% confidence contour. This source
is reported at the 3.9$\sigma$ significance level in the Second EGRET catalog
(Thomson et al. \markcite{Thomp95}1995), but is too weak to be listed as a
catalogued EGRET source. 

The blazar PKS 0208-512 was detected in analysis of Comptel data by Blom et al.
(\markcite{Blom95}1995).  The strongest signal from PKS 0208-512 was obtained
by combining data from 8-13 May 1993 and 3-14 June 1993, yielding a flux in the
1-3 MeV band of $4.1(\pm 0.7)\times 10^{-4}$ photon cm$^{-2}$ s$^{-1}$ and
upper limits at lower and higher energies in the Comptel energy range. 
Contemporary observations with EGRET yield a strong $> 100$ MeV detection of
the source 2EG J0210-5051, with PKS 0208-512 located within the 95\% confidence
contour of this source. The peak in the $\nu F_\nu$ spectrum of PKS 0208-512
occurs at MeV energies, supporting its classification as an MeV blazar. 
Williams et al. (\markcite{Willi95}1995) also report the detection of the
gamma-ray source GRO J1753+57 with a spectrum similar to that of the other two
MeV blazars.  Lack of association of this object with a flat radio spectrum
quasar makes this object's classification uncertain, but one should keep in
mind that blazars are highly variable at all wavelengths, so that
contemporaneous multiwavelength campaigns could be necessary for proper source
identification. 

The peaked emission from MeV blazars has been interpreted as Doppler boosted
$e^+e^-$ annihilation radiation produced in collimated jets of plasma ejected
from a supermassive black hole (Roland \& Hermsen \markcite{Rolan95}1995;
Henri, Pelletier, \&  Roland \markcite{Henri93}1993). The possible importance
of $e^+e^-$ processes and $\gamma$-$\gamma$ opacity effects has been considered
by Blandford \& Levinson (\markcite{Bland94}1994), B\"ottcher, Mause, \&
Schlickeiser (\markcite{B\"ottc97}1997), and Marcowith, Henri, \& Pelletier
(\markcite{Marco95}1995). No attempt has been made, however, to account for the
significant spectral differences between MeV blazars and the $> 50$
$\gamma$-ray blazars detected by EGRET (e.g., von Montigny et al.
\markcite{Monti95}1995), which show strong high-energy radiation extending to
GeV and, in a few cases (e.g., Macomb et al. \markcite{Macom96}1996; Schubnell
et al. \markcite{Schub96}1996), TeV energies. 

In this paper we show that the beaming pattern of thermal Doppler-boosted
$e^+e^-$ annihilation radiation is much broader than the beaming pattern
produced by nonthermal jet electrons which Compton scatter photons produced
outside the jet, whose beaming pattern was recently derived (Dermer
\markcite{Dermer95}1995; Dermer, Sturner, \& Schlickeiser
\markcite{Derme97}1997). We call this latter process external Compton
scattering (ECS) and distinguish it from synchrotron self-Compton (SSC)
scattering, which has also been proposed as the radiation process yielding
blazar gamma radiation (see Marscher \& Travis \markcite{Marsc96}1996 and
references therein).  If the high-energy emission in blazars is primarily
produced by ECS and, moreover, the annihilation luminosity from thermal
$e^+e^-$ pairs is significant, then a class of objects with properties similar
to the MeV blazars is a natural consequence of orientation effects when viewing
at moderate angles with respect to the axis of the outflowing jet. In addition,
we show in Appendix A that if the peaks of MeV blazars are indeed due to
Doppler-boosted thermal annihilation radiation, then  gamma-ray observations to
measure Doppler factors $\cal D$ coordinated with  VLBI radio observations to
measure transverse angular speeds $\mu$ of outflowing radio blobs in two or
more episodes is sufficient to determine the Hubble constant $H_0$. 

\section{Spectra and Angular Distribution of Jet and Accretion-Disk Radiation}

We calculate the received spectral flux and radiation pattern from a system
consisting of an accretion disk and a relativistic collimated outflow of
thermal and nonthermal $e^+e^-$ pairs. We do not attempt to derive the particle
distributions self-consistently from first principles. Evolution of nonthermal
particles due to radiative losses in outflowing plasma blobs has been treated
in detail by Dermer \& Schlickeiser \markcite{Derme93}(1993) and Dermer
et al. \markcite{Dermer97}(1997), and the formation of a
thermalized pair plasma in jets has been considered by B\"ottcher \&
Schlickeiser \markcite{B\"ottc96}(1996). The emphasis here is rather on the
angle-dependence of the received fluxes from the following three spectral
components: (1) annihilation radiation from hot thermal pairs in the
relativistic outflow;  (2) inverse Compton emission from nonthermal pairs in
the outflow; and (3) a quasi-isotropic thermal Comptonization spectrum from hot
optically thin accretion plasma located near the central supermassive black
hole.   The central assumption in calculating the jet spectra, as discussed by
Dermer \& Schlickeiser \markcite{Derme93}(1993), is that the particle
distribution function is isotropic in the comoving blob frame. 

We derive the received spectrum of thermal annihilation radiation.  The
observed flux density $S(\epsilon;\,\mu,\,\phi)$ (erg cm$^{-2}$ s$^{-1}$
$\epsilon^{-1}$) from a blob moving with bulk velocity $\beta c$ is related to
the emissivity in the comoving frame $j^*(\epsilon^*,\,\mu^*,\,\phi^*)$ (erg
cm$^{-3}$ s$^{-1}$ $\epsilon^{-1}$ sr$^{-1}$) through the relation 
\begin{eqnarray}
S(\epsilon;\,\mu,\,\phi)=\frac{{\cal D}^3(1+z)V^*}{d_{\rm L}^2}\,
j^*\left[\frac{\epsilon(1+z)}{{\cal D}},\,\frac{\mu-\beta}{1-\beta\mu},\,
\phi\right]
\end{eqnarray}
(e.g. Begelman, Blandford, \& Rees \markcite{Begel84}1984), where
$\epsilon=h\nu/m_ec^2$ is the dimensionless photon energy in the  stationary
frame and $V^*$ is the volume of the blob in the comoving frame.  (Hereafter
starred quantities refer to the comoving frame, unstarred  quantities to the
stationary frame.) The direction from the source to the  observer is
$(\mu,\,\phi)$, with $\mu$ representing the cosine of the angle between  the
jet axis and the line of sight, and $\phi$ representing the azimuthal angle. 
The Doppler factor ${\cal D}\equiv[\Gamma(1-\beta\mu)]^{-1}$, where  the bulk
Lorentz factor $\Gamma=(1-\beta^2)^{-1/2}$, the speed of the blob is $\beta c$,
and the luminosity distance $d_{\rm L}=2c[1+z-(1+z)^{1/2}]/H_0$ for a
$8\pi G \rho/3H_0^2=1$, $\Lambda=0$ cosmology 
(Weinberg \markcite{Weinb72}1972).  The
Hubble constant $H_0 = 50h$ km s$^{-1}$ Mpc$^{-1}$. 

We assume that there is a population of thermal pairs with dimensionless
temperature $\Theta_a\equiv kT/m_ec^2$ in the outflow. The annihilation
emissivity is isotropic in the comoving frame and is given by the expression 
\begin{eqnarray} 
j_a^*(\epsilon^*,\mu^*,\phi^*;\Theta_a)\; &=& 
\;{L_a^*\epsilon^*\over 4\pi V^*}
\exp\left\{-\frac{1}{\Theta_a}
\left[\epsilon^*+\left(\frac{1}{2\epsilon^*}\right)\right]\right\} \nonumber\\
&&\quad \cdot {\int_1^\infty 
d\gamma_r\,(\gamma_r-1)\,\exp\left[-\gamma_r/(2\Theta_a\epsilon^*)\right]
\,\sigma_a(\gamma_r)\over
\int_1^\infty
d\gamma_r\,(\gamma_r^2-1)\,
K_2\left[\frac{1}{\Theta_a}\sqrt{2(\gamma_r+1)}\right]
\,\sigma_a(\gamma_r)}.
\end{eqnarray}
(Svensson \markcite{Svens83}1983; Dermer \markcite{Derme84}1984), where 
$L_a^*$ is the luminosity of annihilation photons in the comoving frame
and
$K_2$ is the modified Bessel function of the second kind. The integration
variable $\gamma_r={\cal U}_+ \cdot {\cal U}_-$ is the invariant relative
Lorentz factor formed from the $e^+$ and $e^-$ four velocities. The  cross
section for annihilation, expressed in terms of $\gamma_r$, is given by 
\begin{eqnarray}
\sigma_a(\gamma_r)&=&\frac{\pi r_e^2}{\gamma_r+1}\biggl\{\biggl(
\frac{\gamma_r^2+4\gamma_r+1}{\gamma_r^2-1}\biggr) \nonumber \\
&& \ln[\gamma_r+(\gamma_r^2-1)^{1/2}]-\frac{\gamma_r+3}
{(\gamma_r^2-1)^{1/2}}\biggr\}
\end{eqnarray}
(e.g. Jauch \& Rohrlich \markcite{Jauch76}1976), where
$r_e=e^2/m_ec^2=2.82\times10^{-13}$ cm is the classical radius of the 
electron. 

Combining equations (1), (2), and (3)  we obtain the annihilation
flux density $S_a(\epsilon;\,\mu,\phi)$. The  beaming pattern for thermal
annihilation radiation is not obvious from the resulting formula. To derive the
beaming pattern we approximate the annihilation emissivity as a delta function
centered at unity, which is a reasonable approximation when $\Theta_a\ll 1$. 
Thus 
\begin{equation}
j_a^*(\epsilon^*)\simeq\frac{L_a^*}{4\pi V^*}\,
\delta(\epsilon^*-1).
\end{equation}
Inserting this into equation (1) and integrating over $\epsilon$, we find that 
the annihilation flux density integrated over photon energy has a beaming 
pattern
$S_a\propto {\cal D}^4$.  

Electron-positron thermal bremsstrahlung radiation will always accompany the
annihilation radiation.  At temperatures $\theta_{\rm a} \gg 1$, the luminosity
of $e^+$-$e^-$ bremsstrahlung can dominate  e$^+$-e$^-$ annihilation radiation
(see Skibo et al. \markcite{Skibo95}1995).  For the temperatures we use in this
paper, however, the bremsstrahlung flux is less than the  fluxes of the other
components, and is not included in the fits. 

In addition to the annihilation radiation, we assume that there is an isotropic
population of nonthermal pairs in the comoving frame with an energy
spectrum given by a broken  power law, so that
\begin{equation}
n_{\rm nth}(\gamma,\mu,\phi)= \frac{k}{4\pi}
\cases{ \left(\gamma/\gamma_b\right)^{-p_1}
&$\gamma_1\le\gamma\le\gamma_b$\cr
\left(\gamma/ \gamma_b\right)^{-p_2}
&$\gamma_b\le\gamma\le\gamma_2$}\; ,
\end{equation}
where $p_2 = p_1 + 1$ and $p_1 \gtrsim 2$. Such a time-averaged particle
spectrum might arise if shocks are present and radiative cooling is efficient
for pairs in the high energy protion of the spectrum (Dermer \& Schlickeiser
1993; Sikora, Begelman, \& Rees \markcite{Sikor94}1994). We consider a
situation where the particles scatter external ambient soft photons which are
isotropic in the frame through which the radiating plasma moves (Sikora et al.
\markcite{Sikor94}1994).  If the photon energy is
$\epsilon_0^*$, then all scattering will take place in the Thomson regime if
$\Gamma\gamma_2\epsilon_0^* \ll 1$. If the scattering takes place in the
Thomson regime, then the approximation of Reynolds (1982) can be used to derive
(see Dermer \markcite{Derme95}1995; Dermer et al. \markcite{Derme97}1997) 
the observed scattered flux density, given by 
\begin{equation}
S_c(\epsilon;\,\mu,\phi) = \frac{L_{\rm C}^* A \gamma_b^{p_i}}{d_{\rm L}^2}
\left(\frac{1+\mu}{1+\beta}\right)^{1+\alpha_i}
{\cal D}^{4+2\alpha_i}(1+z)^{1-\alpha_i}
\left({\epsilon\over \epsilon_0^*}\right)^{-\alpha_i},
\end{equation}
for $i=1,2$, where $\alpha_i=(p_i-1)/2$. When $i=1$, 
\begin{equation}
\gamma_1^2 \lesssim {\epsilon(1+z)(1+\beta)\over{\cal D}^2\epsilon_0^* (1+\mu)}
\lesssim \gamma_b^2,
\end{equation}
and when $i=2$, 
\begin{equation}
\gamma_b^2 \lesssim {\epsilon(1+z)(1+\beta)\over {\cal D}^2 \epsilon_0^*
(1+\mu)} \lesssim \gamma_2^2.
\end{equation}
The term $L_{\rm C}^*$ in equation (6) represents the luminosity of
Compton-scattered radiation measured in the comoving plasma frame.  The factor
$A$ in the normalization of expression (6) depends only on kinematic quanties
through the expression 
\begin{equation}
A=\frac{3}{16\pi\Gamma^2\epsilon_0^*}\left[\frac{\gamma_b^{p_1}
(\gamma_b^{2-2\alpha_1}-\gamma_1^{2
-2\alpha_1})} {1-\alpha_1}+
\frac{\gamma_b^{p_2}(\gamma_2^{2-2\alpha_2}-\gamma_b^{2-2\alpha_2})}
{1-\alpha_2} \right]^{-1}. 
\end{equation}

The beaming pattern in equation (6) goes as ${\cal
D}^{4+2\alpha}(1+\mu)^{1+\alpha}\sim {\cal D}^{4+2\alpha}$ for scattered 
radiation. If $p_1 \simeq 2$, then the spectral index $\alpha\simeq 1/2$ at
energies below the break energy, and $S_c(\epsilon)\propto {\cal
D}^5(1+\mu)^{3/2}$. Thus the scattered radiation is more highly beamed than the
annihilation radiation, where $S_a\propto {\cal D}^4$.  We note that if the
emission is made by a radiation process which produces a photon source which is
isotropic in the comoving frame, then $S(\epsilon)\propto {\cal D}^{
3+\alpha}$. Such beaming patterns result from isotropically distributed
electrons radiating synchrotron and synchrotron self-Compton emission in a
spherical plasma blob (e.g., Marscher 1987).  This gives a beaming pattern
comparable to the annihilation radiation, and when such emission components
dominate, we expect the ratios of the flux densities of synchrotron or
synchrotron self-Compton and annihilation radiation to vary only weakly with
observing angle. 

The third spectral component we consider is the accretion disk radiation. If
optically thin, this component is isotropic in the stationary frame and would
only display an angular dependence if obscuring clouds or a dusty torus shadow
the radiation.  Such emission could arise from Comptonization of soft accretion
disk photons by a hot disk corona (e.g., Haardt \& Maraschi 1993). For 
simplicity, we model this component by an exponentially truncated power law. 
The observed flux density is then given by 
\begin{equation}
S_d(\epsilon)=\frac{L_{\rm d}(1+z)^{1-\alpha_d}}{4\pi d_{\rm L}^2}\;
\frac{\epsilon^{-\alpha_d}\,\exp\left[-(1+z)\epsilon/\Theta_d\right]}
{\int_{\epsilon_c}^\infty d\epsilon\,\epsilon^{-\alpha_d}\,
\exp\left(-\epsilon /\Theta_d\right)},
\end{equation}
where the disk luminosity is denoted by $L_{\rm d}$. 

In Fig. 1 we show spectra for various viewing angles, choosing $\Gamma=5$. The
pair temperature in the outflow is fixed at $\Theta_a=0.5$ and the parameters
of the nonthermal pair distribution are $\gamma_1=1$,  $\gamma_b=10^3$,
$\gamma_2=10^6$, $p_1=2.1$  and $p_2=3.1$. We set $\alpha_d=0.5$,
$\Theta_d=0.5$, and $\epsilon_c=10^{-3}$ for the disk parameters and set the
redshift $z =1$. The disk luminosity $L_d $ is assigned the value $10^{47}$
ergs s$^{-1}$.  We set  $L_{\rm C}^* = L_{\rm d}/\Gamma$.  The reason for this
choice is that the jet luminosity should normally be less than the disk
luminosity, since the accretion power is supplying the jet luminosity, and the
factor $\Gamma^{-1}$ accounts for the additional power required by the bulk
motion of the jet.  For the annihilation luminosity, we take $L_{\rm a}^* = 0.1
L_{\rm C}^*$. 

As can be seen in Fig. 1, the disk radiation is independent of observing angle
and the annihilation radiation is less beamed than the strongly anisotropic
Compton-scattered radiation. For this simple model, the disk produces X-ray
emission only, the Compton scattered radiation always dominates at gamma-ray
energies above 10 MeV and the annihilation radiation appears as a enhancement
in the MeV range which, for certain observing angles, can dominate the flux at
MeV energies. Hence for small observing angles, the source appears as a
gamma-ray emitting blazar with the luminosity dominated by high-energy
gamma-rays and the source displays properties similar to a large fraction of
the $> 50$ gamma-ray blazars detected with EGRET (see, e.g., von Montigny et
al. \markcite{vonmo95}1995). At angles slightly off axis the high-energy
radiation flux decreases and the bulk of the high-energy gamma-ray luminosity
comes out in the MeV range. The annihilation feature can dominate at moderate
observing angles. At larger angles, the disk radiation dominates the observed
flux, although scattered jet radiation could make a substantial contribution to
the flux at MeV energies (Skibo, Dermer, \& Kinzer \markcite{Skibo94}1994). 
Note also that radio loud AGN could exhibit time-variable hardenings and
features at MeV energies due to the electron-positron component, as suggested
in the $\theta = 30^\circ$ panel in Fig. 1. 

We use this model to fit the Comptel data for the blazars PKS 0208-512 at
redshift $z=1.003$ and PKS 0506-612 at $z=1.093$ (Bloemen et al.
\markcite{Bloem95}1995; Blom et al. \markcite{Blom95}1995). In Figs. 2 and 3 we
show the contemporaneous gamma-ray observations of these sources made with the
Comptel and EGRET instruments on {\it CGRO}. The X-ray points are
noncontemporaneous observations made with {\it ROSAT} (Brinkman et al.
\markcite{Brink94}1994). In these fits, we fix the relative luminosities of the
various spectral components to satisfy $L^*_a=L^*_{\rm C}=\Gamma^{-1}L_d$. For
the case of PKS 0208-512, the bulk Lorentz factor $\Gamma=3$, the inclination
angle $\theta=18^\circ$ and the luminosity $L_d=10^{47}$ erg s$^{-1}$. For PKS
0506-612, we let $\Gamma=11$, $\theta=9^\circ$ and $L_d=3\times10^{46}$ erg
s$^{-1}$. All other parameters are the same as used for Fig. 1.  The parameters
used in the fits are not unique, but are typical for gamma-ray blazars.  Thus
if jets are composed of electrons and positrons which thermally annihilate,
then the existence of a class of MeV blazars could be a straightforward 
consequence of the different beaming patterns of the radiations.

\section{Discussion}

In this paper, we have considered implications of the suggestion that the
features in MeV blazars arise from thermal annihilation radiation in a
relativistic jet, as suggested by previous authors (see Section 1).  The
existence of a class of MeV blazars can be understood by orientation effects if
the beaming patterns of the nonthermal radiation is different from that of the
annihilation radiation.  If the gamma rays are produced by nonthermal jet
electrons scattering photons produced external to the jet, then the
annihilation radiation has a broader beaming pattern.  This could account for
the existence of the class of MeV blazars, and would be consistent with
unification scenarios for radio-emitting AGNs (see, e.g.,  Urry \& Padovani
\markcite{Urry95}1995 for a recent review). Moreover, if this interpretation is
correct, then we predict that radio galaxies will exhibit time variable
hardenings and features at MeV energies which would correlate with increasing
core dominance. 

Gamma-ray observations of line features and contemporaneous VLBI measurements
of transverse angular speeds from relativistic plasma outflows provide a method
to determine the Hubble constant. The radio observations furnish the  apparent
speeds, and the gamma-ray observations give the Doppler and cosmologically
shifted energy of the annihilation line. As noted earlier by  Roland \& Hermsen
\markcite{Rolan95}(1995), measurements of $\cal D$ and the apparent
superluminal speed can be used to determine the angle of the jet with respect
to the observer and the bulk Lorentz factor $\Gamma$ of the outflowing plasma.
As shown in Appendix A, multiple contemporaneous observations of the shifted
line energies and transverse angular speeds provide a method to determine the
Hubble constant. This is an extension of
the work of Schlickeiser \& Dermer (\markcite{Schli95}1995), who proposed a
model-dependent method for measuring $\cal D$ from the broadband spectral
energy distribution of blazars. If the MeV features can be conclusively
identified with annihilation line signatures, then the procedure proposed here
will provide a more definitive method for determining the Hubble constant from
coordinated radio and gamma ray observations. 

Arbitrary spectral lines emitted from the outflowing plasma jets can also be
utilized by this method. The identification of spectral line features with a
specific outflowing plasma blob is indicated at gamma-ray energies by time
variability. Observations (Wehrle \markcite{Wehrl96}1996) show that the
emergence of a radio emitting feature in high-resolution VLBI maps is preceded
by a gamma-ray flare.   A better method would be through milliarcsecond imaging
of spectral lines, which is presently only feasible at radio and
optical wavelengths; high resolution optical imaging requires, however, 
bright sources of apparent magnitudes $m \lesssim 9$. 

If it is established that the peaks of MeV blazars are
a consequence of electron-positron plasma jets, then such a result would have important
implications for processes which power jets in radio-loud AGNs, requiring energization of
particles in the compact environment near the supermassive black hole.  Additional measurements
with Comptel on {\it CGRO}, and observations with more sensitive soft gamma-ray telescopes, such
as the {\it INTEGRAL} telescope scheduled for launch early in the next millenium, could resolve
the question of the composition of jets in AGNs and provide a test of this model.

\begin{acknowledgements}

The work of C.D. was partially supported by NASA DPR S-57770-F. R.S. 
acknowledges partial support by DARA (50 OR 9406 3). J.S. and C.D. acknowledge
support from Office of Naval Research.

\end{acknowledgements}

\appendix
\section{A Method for Determining the Hubble Constant}

Here we show how gamma-ray observations may be used in conjunction with VLBI
rdio observations to measure the Hubble constant. Consider a source at redshift
$z$ which ejects two $e^+e^-$ pair plasma blobs  with velocities $\beta_1c$ and
$\beta_2c$ at different times in the same direction $\theta$ measured with
respect to the line of sight. Let $\delta_1$ and $\delta_2$ be the blueshifts
of the  annihilation line emitted by these blobs measured with $\gamma$-ray 
observations. We have 
\begin{eqnarray}
\delta_1=[\Gamma_1(1-\beta_1\cos\theta)(1+z)]^{-1}, \\
\delta_2=[\Gamma_2(1-\beta_2\cos\theta)(1+z)]^{-1},
\end{eqnarray}
where $\Gamma_i=(1-\beta_i^2)^{-\frac{1}{2}}$, $i=1,2$. Eqs. (A1) and (A2)
assume that the peak of the annihilation radiation occurs at m$_{\rm e}$c$^2$,
whereas the peak of thermal annihilation radiation is blueshifted (Ramaty \&
Meszaros \markcite{Ramat81}1981).  The blueshifting of the peak of the
radiation can be accounted for by determining the temperature through the width
of the emission feature.  If this effect is properly taken into account, then
$\delta_1$ and $\delta_2$ can be precisely determined. 

Let $\mu_1$ and $\mu_2$ be the angular speeds of the two blobs determined from
radio observations. The apparent speeds of the blobs in the plane of the sky
are 
\begin{eqnarray}
\frac{d_M\mu_1}{c}&=&\frac{\beta_1\sin\theta}{1-\beta_1\cos\theta}
 = (1+z)\beta_1\Gamma_1\delta_1\sin\theta,
\\
\frac{d_M\mu_2}{c}&=&\frac{\beta_2\sin\theta}{1-\beta_2\cos\theta}
= (1+z)\beta_2\Gamma_2\delta_2\sin\theta.
\end{eqnarray}
where $d_M$ is the proper motion distance. Setting the cosmological
deceleration parameter $q_0=1/2$ the proper motion distance is a function of
the cosmological redshift $z$ and the Hubble constant $H_0$ (Weinberg 1972)
through the expression
\begin{equation}
d_M=\frac{2c}{H_0}\left[1-(1+z)^{-\frac{1}{2}}\right].
\end{equation}

These five equations express the five unknown quanties $d_M$, $\theta$,
$\beta_1$, $\beta_2$ and $H_0$ in terms of the measured quantities $\mu_1$,
$\mu_2$, $\delta_1$, $\delta_2$ and $z$. Solving equations (A1) and (A2) for
$\cos\theta$ and equations (A3) and (A4) for $\sin\theta$, and equating the
ratio $\tan\theta$ to eliminate $\theta$, we obtain 
\begin{equation}
\Gamma_2=\frac{1}{\delta_2(1+z)}+
\left(\frac{\mu_2\delta_1}{\mu_1\delta_2}\right)
\left[\Gamma_1-\frac{1}{\delta_1(1+z)}\right].
\end{equation}
Dividing equation (A3) by (A4) leads to 
\begin{equation}
\Gamma_1^2=1+\left(\frac{\mu_1\delta_2}{\mu_2\delta_1}\right)^2
\;(\Gamma_2^2-1).
\end{equation}
Substituting (16) into (17) we arrive at the following expression for 
$\Gamma_1$:
\begin{equation}
\Gamma_1=\left[\frac{2}{\delta_1(1+z)}\left(1-\frac{\mu_1}{\mu_2}\right)\right]^
{-1}
\left\{ 1-\left(\frac{\mu_1\delta_2}{\mu_2\delta_1}\right)^2+
\left[\frac{\left(1-\frac{\mu_1}{\mu_2}\right)}{\delta_1(1+z)}\right]^2\right\}.
\end{equation}
From equation (A1) we then determine $\theta$, given by
\begin{equation}
\theta=\cos^{-1}\{\beta_1^{-1}-[\beta_1\Gamma_1\delta_1(1+z)]^{-1}\}.
\end{equation}
Using equation (A3) for the proper motion, we obtain
\begin{equation}
d_M=\frac{c(\Gamma_1^2-1)^{1/2}}{\mu_1}(1+z)\delta_1\sin\theta,
\end{equation}
and finally from equation (A5) we find the Hubble constant, given by
\begin{equation}
H_0=\frac{2c}{d_M}\left[1-(1+z)^{-1/2}\right].
\end{equation}

\eject

\eject

\begin{figure}
\centerline{\psfig{figure=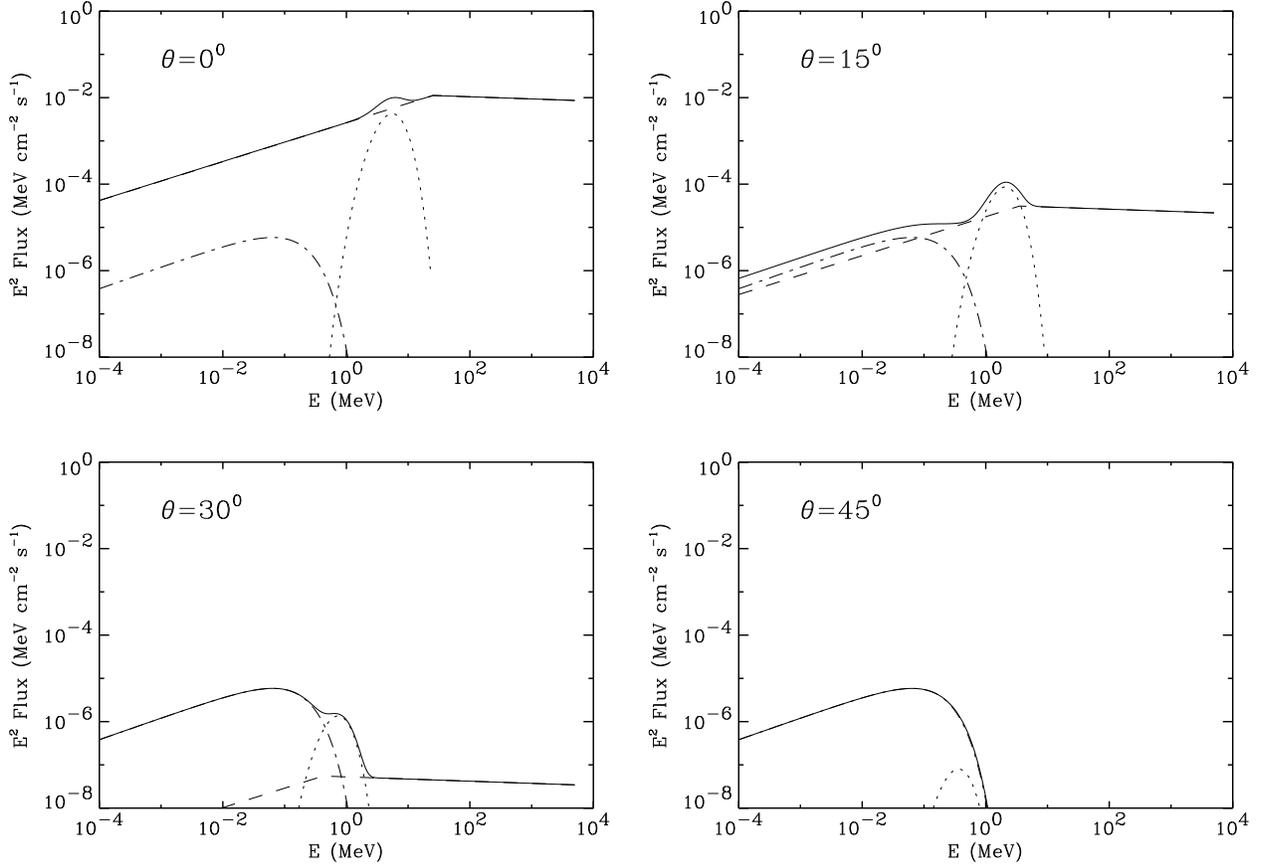,height=12 cm,width=17 cm}}
\caption{ \em
Spectra for various viewing angles and bulk Lorentz factor
$\Gamma = 5$. The dashed curves represent the observed flux produced by inverse
Compton scattering of isotropic soft photons by nonthermal pairs in the
relativistic outflowing plasma jet, the dotted curves are the observed fluxes
of thermal pair annihilation radiation produced in the jet, and the dot-dashed
curves are the optically thin isotropic emission produced by an accretion disk
corona.  The solid curves give the total fluxes.  See text for parameter
values. 
}
\end{figure}

\begin{figure}
\centerline{\psfig{figure=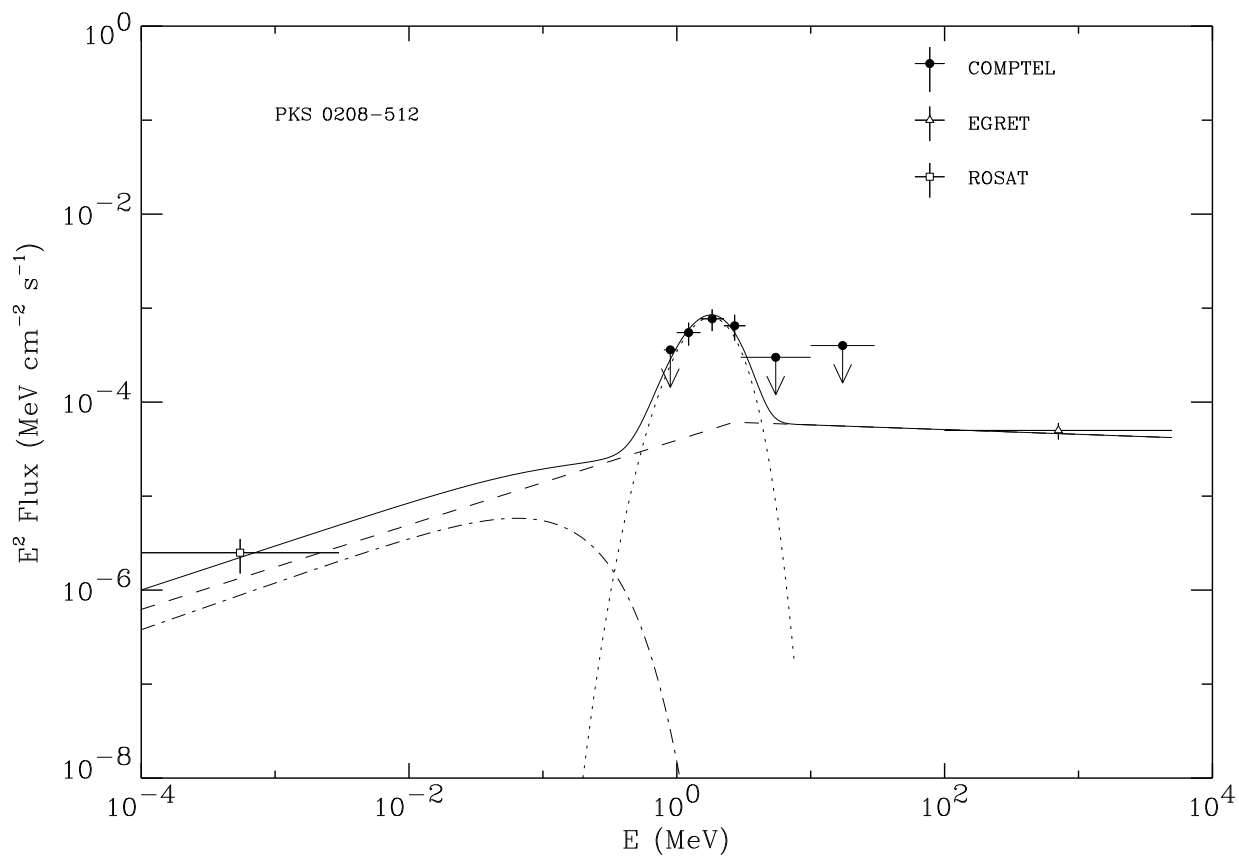,height=12 cm,width=17 cm}}
\caption{ \em
High-energy spectrum of PKS 0208-512. The individual 
components are specified by the curve types as in Fig. 1. The parameters
are  
$\alpha_d=0.5$, $\Theta_a=\Theta_d=0.5$, $\epsilon_c=10^{-3}$,
$\gamma_1=1$, $\gamma_2=10^6$, $\gamma_b=10^3$, $p_1=2.1$, $p_2=3.1$,
$\Gamma=3$ and $\theta=18^\circ$.
}
\end{figure}

\begin{figure}
\centerline{\psfig{figure=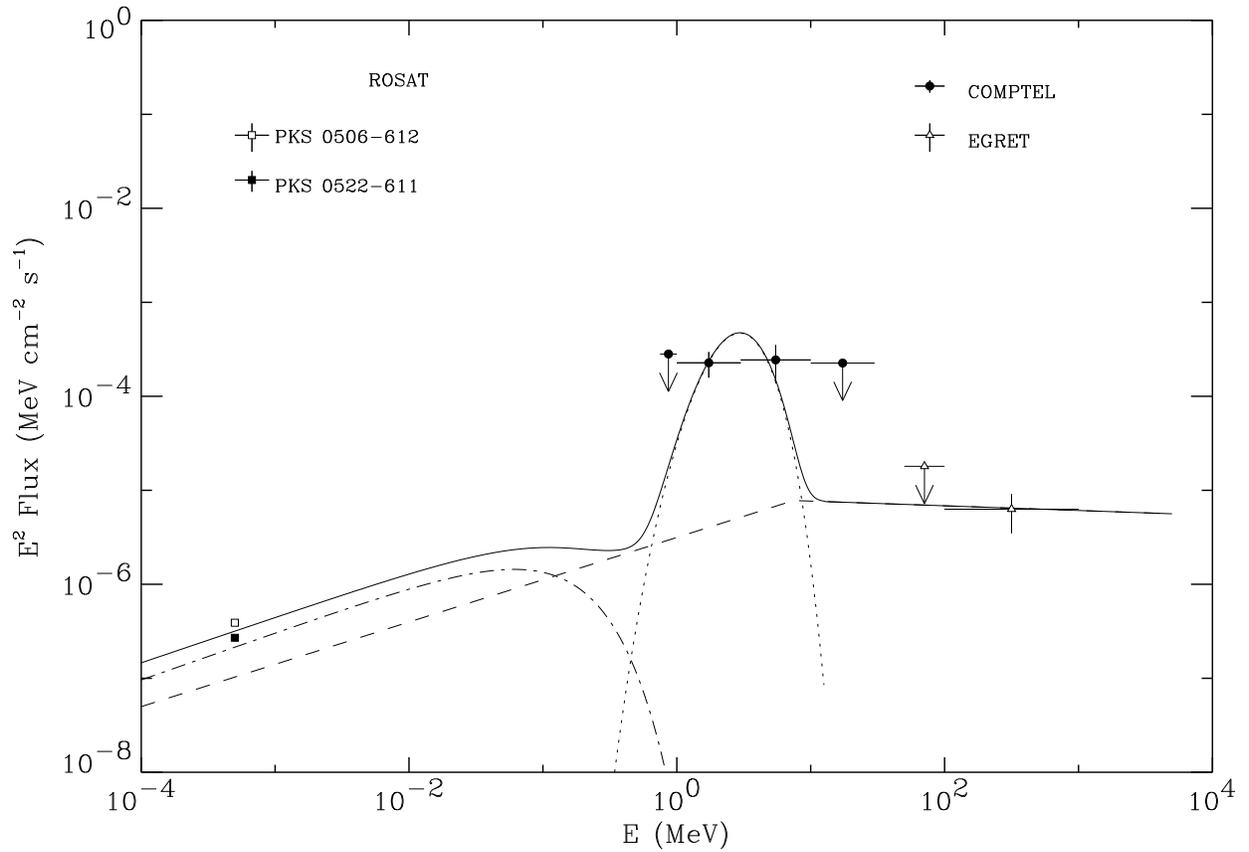,height=12 cm,width=17 cm}}
\caption{ \em
High-energy spectrum of PKS 0506-612. The individual 
components are specified by the curve types as in Fig. 1. The parameters
are the same as in Figure 2 except 
$\Gamma=11$ and $\theta=9^\circ$.
}
\end{figure}

\end{document}